\documentclass[aps,amsmath,amssymb,prb,twocolumn,showpacs,superscriptaddress]{revtex4}
\usepackage{bm}
\usepackage{epsfig}
\usepackage[usenames]{color}
\usepackage{graphicx}

\newcommand{\inel}{{\rm in}}
\newcommand{\ph}{{\rm ph}}
\newcommand{\T}{{\mathfrak{T}}}
\newcommand{\dt}{{\tau}}
\newcommand{\V}{{\mathcal{V}}}

\renewcommand{\paragraph}[1]{\textit{#1.---} } %PRL paragraph style: italic & .--- added

\begin{document}

\title{Single grain heating due to inelastic cotunneling}

\author{Andreas~Glatz}
\affiliation{Materials Science Division, Argonne National Laboratory, Argonne, Illinois 60439, USA}

\author{I.~S.~Beloborodov}
\affiliation{Department of Physics and Astronomy, California State University Northridge, Northridge, CA 91330, USA}

\date{\today}

\begin{abstract}
We study heating effects of a single metallic quantum dot weakly coupled to two leads.
The dominant mechanism for heating at low temperatures is due to inelastic electron cotunneling processes.
We calculate the grain temperature profile as a function of grain parameters, bias voltage, and time and show that for nanoscale size grains the heating effects are pronounced and easily measurable in experiments.
\end{abstract}

\maketitle
%%%%%%%%%%%%%%%%%%%%%%%%%%%%%%%%%%%%%%%%%%%%%%%%%%%%%%%%%%%%%%%%%%

\begin{figure}[tbh]
\includegraphics[width=0.5\columnwidth]{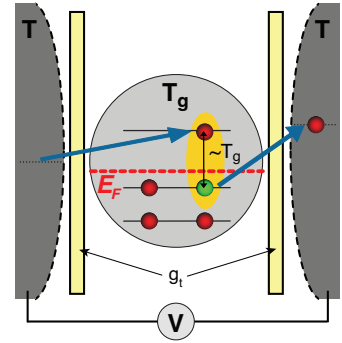}
\caption{(Color online) Sketch of a single grain weakly coupled to two leads, $g_t \ll 1$. 
Temperatures $T$ and $T_g$ refer to the leads and grain temperatures, respectively. 
Arrows indicate the inelastic cotunneling process through a single grain with a typical energy of the electron-hole pair $T_g$. 
The dashed line corresponds to the Fermi level in the grain. 
Here only the electron propagation process responsible for the heating of the grain is shown ($\sigma_\inel$). 
The heat is removed from the grain by a similar process but with the electron starting and ending in the same lead (inelastic cotunneling loop, $\kappa_\inel$) and by phonons ($\kappa_{ph}$).}
\label{fig.model}
\end{figure}

Great efforts in contemporary materials science research focus on the understanding of thermal properties of nanoscale devices~\cite{RMP06,Pekola07,Saira07}.
The simplest possible nanodevice consists of a single grain coupled to two leads (see Fig.~\ref{fig.model}).
The transport properties through this system are well understood~\cite{Averin92,Glazman}. 
However, much less is known about heating effects in this system.
The understanding of heating effects on properties of nanodevices is especially important for practical applications.
Indeed, recent experimental research~\cite{Heath08,Shi03,Hoffmann07} has focused on the application of the thermal properties of low-dimensional devices for efficient power generation.
This defines an urgent quest for a quantitative description of heating effects on properties of nanodevices.

Indeed, the effect of heating is very important for low-temperature nano-thermoelectric devices.
In particular, next-generation devices made from nano-granular materials can be influenced by internal heating, and it is important to know the device's internal temperature.
Granular thermoelectric materials have the advantage that one can control the system parameters and therefore the device properties.
A measure for the efficiency of a thermoelectric material is the dimensionless {\it figure of merit}, $ZT$, where $T$ is the temperature, which depends on the thermopower or Seebeck coefficient, $S$ and the electric, $\sigma$ and thermal, $\kappa$ conductivities, $ZT = S^2\sigma T/\kappa$.~\cite{Rowe}
One has to take into account the fact that all three kinetic coefficients, $\sigma$, $\kappa$, and $S$, depend on the {\it internal} device temperature.~\cite{Majumdar,venka01,harman02}
Recently we investigated the properties of metallic granular metals in the metallic~\cite{glatz+prb09b} and weakly coupled regimes~\cite{Glatz09} which are relevant for applications as low-temperature thermocouples.
Here we will answer the question to what extend heating effects influence the properties of a single quantum dot -- the building block of granular metals.

In this paper we investigate the effect of inelastic cotunneling on heating of a grain weakly coupled to two leads
(see Fig.~\ref{fig.model}).
We calculate the grain temperature profile as a function of grain parameters,
bias voltage, and time.
For nanoscale size grains it is shown that the heating effects are pronounced and easily
measurable in experiments.
Here we consider the case of an isolated grain in the sense that it is not thermally coupled to an environment other than the leads, e.g. the grain is not coupled to a substrate.

A single grain is characterized by two energy scales: (i) the mean energy level spacing $\delta$ and (ii) the charging energy $E_c $. 
We concentrate on the case of metallic grains which satisfy the inequality $E_c \gg \delta$.

In addition to the above energy scales the system under consideration is characterized by the tunneling conductance $g_t$. 
In this paper we concentrate on the regime of Coulomb blockade corresponding to a weak coupling between the grain and the leads, $g_t \ll 1$. 
Our considerations are valid for temperatures $ \delta < T < E_c$.

\paragraph{Transport mechanisms} 
The only transport mechanism contributing to grain heating is due to inelastic cotunneling.
Cotunneling allows simultaneous charge transport through several junctions by means of cooperative
electron motion. 
Single cotunneling (Fig.~\ref{fig.model}), introduced in Ref.~[\onlinecite{Averin}] provides a conduction channel at low applied biases, where otherwise the Coulomb blockade arising from electron-electron repulsion would suppress the current flow. 
The essence of a cotunneling process is that an electron tunnels via virtual states in intermediate granules thus bypassing the huge Coulomb barrier.
This can be visualized as coherent superposition of two events: tunneling of the electron into a granule and the simultaneous escape of another electron from the same granule. 
There are two distinct mechanisms of cotunneling processes, elastic- and inelastic cotunneling.
Elastic cotunneling means that the electron that leaves the dot has the same energy as the incoming one. 
In the event of inelastic cotunneling, the electron coming out of the dot has a different energy than the entering electron. 
This energy difference is absorbed by an electron-hole excitation in the dot, which is left behind in the course of the inelastic cotunneling process, Fig.~\ref{fig.model}. 
Below we concentrate on inelastic cotunneling because only this transport mechanism contributes to heating effects.
In particular, elastic cotunneling and sequential tunneling do not create electron-hole pairs in the grain.

In the following we consider the electric conductivity due to inelastic processes, which heat the grain, and the electronic and phonon thermal coupling to the leads, which remove excess heat from the grain, for the description of the complete heating effect of a single grain.

\paragraph{Grain heating} The grain heating due to inelastic cotunneling can be described by the following kinetic equation
\begin{equation}
\label{Tg1}
c_v \frac{\partial T_g}{\partial t} = \sigma_\inel \left(\frac{V}{a}\right)^2 - 2 \kappa \frac{T_g - T}{a^2},
\end{equation}
where $c_v = (1/3) T_g \nu$ is the heat capacitance of the grain with $\nu$ being the density of states in the grain; 
$\partial T_g/\partial t$ describes the change of the grain temperature $T_g$ in time $t$; 
$\sigma_\inel$ is the cotunneling conductance (defined below), $V$ being the bias voltage and $a$ the grain size; 
and $\kappa$ is the thermal conductivity, with $\kappa=\kappa_\inel+\kappa_\ph$ where $\kappa_\inel$ is the electron and $\kappa_\ph$ the phonon contribution to the thermal conductivity. 
Equation~(\ref{Tg1}) has a transparent physical meaning:
The energy of the electron-hole pair is released as heat, see also Fig.~\ref{fig.model} whereas the thermal conduction to the leads removes the heat from the grain until a steady state is reached.
The typical energy of electron-hole pairs is of order of the grain temperature, $T_g$, and included in the expressions for $\sigma_\inel$ and $\kappa_\inel$.
The leads are assumed to be a heat sink and stay at ambient temperature $T$ since they are much larger than the grain.
The first term on the r.~h.~s. of Eq.~(\ref{Tg1}) describes the energy released in the grain from an electron-hole pair times the square of the discrete gradient of voltage ($V/a$).
The latter ensures that the result is invariant under a sign change of the voltage $V$. The second term describes the heat removal from the grain driven by the discrete laplacian of the temperature ($2(T_g - T)/a^2$) with its rate $\kappa$, cf. Eq. (23) of Ref.~[\onlinecite{RMP06}].
Our considerations are valid as long as the applied voltage does not break down the Coulomb blockade, i.e. when $eV<E_c$.

Before solving Eq.~(\ref{Tg1}) for temperature $T_g$ vs. time $t$ exactly, we discuss its stationary solution
corresponding to large times, $t \gg t^*$, where $t^*$ is some characteristic time scale defined below.
In other words we first estimate how pronounced the effect of inelastic cotunneling on grain heating is.
For times $t \gg t^*$ the grain temperature reaches some constant steady state value, $T_g  \equiv T_g^*$, i.e. $\partial T_g/\partial t=0$. Using Eq.~(\ref{Tg1}) one obtains
\begin{equation}\label{Tgs1}
\sigma_\inel V^2 = 2 \kappa (T_g^* - T).
\end{equation}
The inelastic cotunneling electric, $\sigma_\inel$, and thermal, $\kappa$, conductivities/conductance of a single metallic grain weakly
coupled to two leads are given by the following expressions~\cite{Glazman,Averin,Basko,Tripathi06,Beloborodov07,Glatz09}
\begin{subequations}
\begin{eqnarray}
\sigma_\inel &=& 2e^2 a^{-1} g_t^2 \frac{T_g^2 + (eV)^2}{E_c^2} \label{sigma}, \\
\kappa_\inel &=& \gamma_e a^{-1} g_t^2 T_g \frac{T_g^2 + (eV)^2}{E_c^2}\label{kappae},\\
\kappa_\ph &=& \gamma_\ph l_\ph^{-1} T_g \left(\frac{T_g}{\Theta_D}\right)^2.\label{kappaph}
\end{eqnarray}
\end{subequations}
In Eq.~(\ref{kappae}) and (\ref{kappaph}) $\gamma_{e/\ph}$ are numerical coefficients ($\gamma_e=32\pi^3/15$ and $\gamma_\ph=8\pi^2/15$, see [\onlinecite{Glatz09}]), $l_\ph$ the phonon mean free path $l_\ph\sim a$,
and $\Theta_D$ the Debye temperature.
Notice, that the temperature entering these expressions is the grain temperature, $T_g$, which determines the number of typically involved internal energy states of the grain\cite{Glazman}.
Substituting these expressions back into Eq.~(\ref{Tg1}) we obtain
the following equation of motion
\begin{equation}
\label{Tg2}
\T_g\frac{\partial \T_g}{\partial\dt}=\alpha\left[\V^2\left(\T_g^2+\V^2\right)-(\T_g^2-\T_g)\left(\gamma\T_g^2+\gamma_e\V^2\right)\right]\,,
\end{equation}
where we introduced the dimensionless grain temperature $\T_g=T_g/T$, time $\dt=t\delta$, and voltage $\V=eV/T$, and used the dimensionless parameters\footnote{In typical situations the parameter $\gamma$ is much larger than $\gamma_e$.}
\begin{subequations}
\begin{eqnarray}
\alpha&=&6 g_t^2\left(\frac{T}{E_c}\right)^2\,,\label{alpha}\\
\gamma&=&\gamma_e+\gamma_\ph  \frac{a}{l_\ph } \left(\frac{E_c}{g_t\Theta_D}\right)^2\,.\label{gamma}
\end{eqnarray}
\end{subequations}

\begin{figure}[tbh]
\includegraphics[width=0.75\columnwidth]{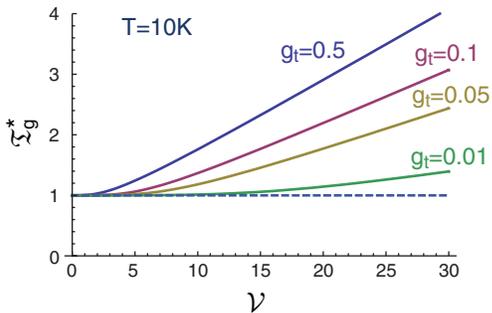}
\caption{(Color online) Voltage dependence of the dimensionless steady state grain temperature
$\T_g^*$ [solution of Eq.~(\ref{Tgs1})] for different values of the tunneling conductance $g_t$, shown next to the graphs. 
All graphs are plotted for the typical parameters given in the text and at temperature $T=10K$. The values for $\V$ are valid for $\V<E_c/T$.}
\label{fig.TgsV}
\end{figure}

Using these definitions, the solution of Eq.~(\ref{Tgs1}) for dimensionless grain temperature in the limit $\T_g^*\gg\V$
is given by
\begin{equation}\label{Tgs2}
\T_g^*=\frac{1}{2}\left(1+\sqrt{1+4\V^2/\gamma}\right)\,,
\end{equation}
which can be simplified for small voltages $\V^2\ll \gamma$ to $\T_g^*=1+\V^2/\gamma$. The complete voltage dependence of $\T_g^*$ is shown in Fig.~\ref{fig.TgsV}.

\paragraph{Time scale and estimates}
We define the typical timescale associated with the heating of the grain due to inelastic cotunneling as the time $t^*$ when the grain temperature $T_g$ reaches the intermediate value $(T_g^*+T)/2$, which can be calculated in the limit $\T_g^*\gg\V$ as
\begin{equation}
\label{dts}
\dt^*=\frac{\gamma}{2\alpha\V^2}\left[\ln\left(\frac{\V^2(\T_g^*+1)^2}{3\V^2+\gamma(1-\T_g^*)}\right)-\frac{\ln\left(3-\T_g^{*-1}\right)}{2\T_g^*-1}
\right]\,,
\end{equation}
with $\T_g^*$ defined in Eq.~(\ref{Tgs2}).
For small voltages, i.e. $\V^2\ll \gamma$, Eq.~(\ref{dts}) can be simplified to $\dt^*=\left(\ln 2+(1-3\ln 2)\V^2/\gamma\right)/(\gamma\alpha)$ meaning that the typical timescale $\tau^*$ is only weakly voltage dependent.

\begin{figure}[tbh]
\includegraphics[width=0.98\columnwidth]{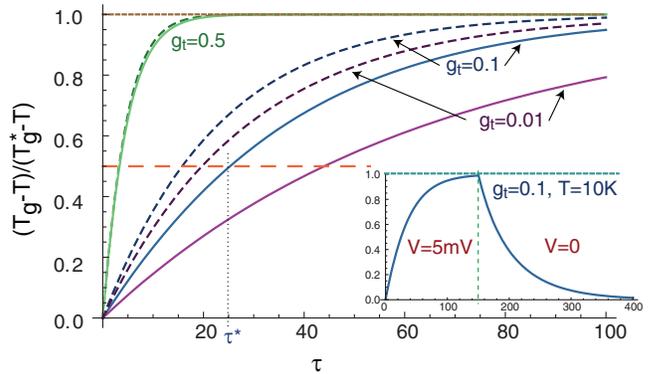}
\caption{(Color online) Plots of the grain temperature $T_g$ vs. dimensionless time $\dt=t\delta$. 
The characteristic scale $T_g^*$ is given by Eq.~(\ref{Tgs1}) and $\dt^*$ approximately by Eq.~(\ref{dts}). 
The y-axis is shifted by $T$ and scaled to $1$ for better comparison. 
The graph shows plots for three different values of the tunneling conductance $g_t$ written next to each group of two curves which represent different values for ambient temperature $T$: $T=10K$ (solid) and $T=15K$ (dashed). 
The inset shows the process of heating and cooling (with the same axis as the main plot): in the left part ($\tau<150$) a finite voltage is applied [$V=5mV$ in Eq.~(\ref{dts})] which is switched off at $\dt=150$ [$V=0$]. }
\label{fig.Tg}
\end{figure}

Next, we will use typical experimental parameters and estimate the heating effect of an isolated grain weakly coupled to two leads with tunneling conductance $g_t=0.1$.
For this we can first estimate the dimensionless parameter $\gamma$ [Eq.~(\ref{gamma})] for a $a=10$nm metallic grain. 
For Aluminum, Nickel, or Iron the Debye temperature and Coulomb energy are $\Theta_D\approx 450$K and $E_c\approx 1600$K, respectively, which results in $\gamma\approx 7\times 10^3$. At a low temperature of $T=10K$ and typical voltages $V\approx 5$mV, i.e. $\V=5.8$, we have for the dimensionless parameter $\alpha=3\times 10^{-6}$. 
Therefore the resulting increase of temperature is about $1K$ and the resulting timescale for this temperature increase for a typical mean level spacing of $1$K is $t^*\approx 0.2$ns (see Figs. \ref{fig.TgsV} \& \ref{fig.Tg}).

\paragraph{Discussions}
The typical timescale $t^*$ for the heating process is, although rather short, still observable with state-of-the-art transport measurements which can have picosecond resolution~\cite{book_fast}.
However, one has to take into account that also sequential and elastic cotunneling can contribute to the total electric current through the grain\footnote{However, sequencial tunneling and elastic cotunneling do not contribute to the heating effects.}.
Compared to the timescale $t^*$, the typical life-time of an electron-hole pair ($\propto 1/T_g^*$) is 2-3 orders of magnitude shorter, such that our assumption of instant heat generation from the electron-hole pairs is well justified. 
Also the heat distribution inside the grain and leads due to phonons happens on a much smaller timescale than $t^*$ such that we can assume that electrons and phonons are in thermal equilibrium within the grain or lead systems.
At this point we note that in general the kinetic Eq.~(\ref{Tg1}) has an additional term proportional to the thermoelectric/Seebeck coefficient. 
However, this term is much smaller -- at least one order of magnitude -- than the electric and thermal conductivities, see Ref.~[\onlinecite{Glatz09}].

Finally, we can also study the single grain system in a steady state for voltages $V>0$ which is switched off to $V=0$ at a certain time $t_0$. 
In this case one should investigate Eq.~(\ref{Tg1}) in the absence of the first term on the r.~h.~s. with the initial condition $T_g(t_0)=T_g^*(V>0)>T$, where $T$ is the lead's temperature. 
The time dependence of the cooling process is shown in the inset of Fig.~\ref{fig.Tg}, where the voltage $V$ is switched off at the dimensionless time $\dt_0=150$. 
The typical timescale of the cooling process is then given by Eq.~(\ref{dts}) with $\V=0$, i.e. $\dt^*=\ln 2/(\gamma\alpha)$ which is longer than for the heating process.

To conclude, we derived the kinetic equation for the heating of a single grain weakly coupled to two leads. 
The heating is governed by inelastic cotunneling processes which create electron-hole pairs in the grain. 
The heat is removed  from the grain by electron and phonon thermal transport processes. Using this equation we obtained the steady state temperature of the grain and the typical timescale which is needed to reach this state.
For a typical metallic grain of $10$nm diameter at ambient temperature $T=10$K the grain temperature rises about $10\%$ for an applied voltage of $5$mV and the related timescale is of order $0.2$ns which is a noticeable important effect for low temperature thermoelectric devices like thermocouples for temperature measurements.

\acknowledgements
A.G. is grateful to N.M. Chtchelkatchev, K.I. Matveev, and V.M. Vinokur for useful discussions.
This work was supported by the U.S. Department of Energy Office of Science under the Contract No. DE-AC02-06CH11357.

\vspace{-0.5cm}

%%%%%%%%%%%%%%%%%%%%%%%%%%%%%%%%%%%%%%%%%%%%%%%%%%%%%%%%%%%%%%%%%%

\end {document}